\\

Title: Express method of gamma-ray analysis of the soil blocks which have been sampled without a disturbance of the turf layer
Authors: E. G. Tertyshnik, S. M. Vakulovsky
This paper presents express method of gamma-analysis of the soil patterns sampled soon after of nuclear accident. The method does not require of the sample preparation and intends for analysis of the soil samples with a non-homogeneous distribution of activity at the depth. Technique of calibration of the detector efficiency is considered, that have been used when soil blocks sampling by means of the non-disturbance method (by rings) after Chernobyl accident were measured.
   Comment:   7 pages, including 2 table
   Subjects:    Nuclear Experiment (nuc-ex), Geophysics (physics.geo-ph)
   Joural reference:    ANRY, 2 (61), 2010, 31-34 (in Russian)

-----------

   Заголовок: Экспрессный метод гамма-спектрометрического анализа почвенных блоков, отобранных без нарушения дернового слоя
   Авторы: Э. Г. Тертышник, С. М. Вакуловский
   Описан экспрессный способ гамма-анализа образцов почвы, отобранных вскоре после радиационной аварии. Способ не требует пробоподготовки и рассчитан для анализа почвенных образцов с неоднородным распределением активности по глубине. Рассмотрена техника калибровки гамма-спектрометра по эффективности регистрации, использованная при анализе проб почвы, отобранных не деструктивным методом (с помощью колец), после аварии на Чернобыльской АЭС.
   Комментарий:    7 страниц, включая 2 таблицы
   Предмет:       Ядерный эксперимент, Геофизика
   Журнальная публ.:    АНРИ, 2 (61), 2010, 31-34
\\

\\

**Экспрессный метод гамма-спектрометрического анализа почвенных блоков, отобранных без нарушения дернового слоя**

Э. Г. Тертышник, С. М. Вакуловский

*ФГБУ «НПО «Тайфун», г.Обнинск, Калужская обл.*

Уровни загрязнения местности радионуклидами после радиационных аварий определяют путём отбора проб почвы и последующего радионуклидного анализа. Известно, что продукты деления, выпавшие на поверхность почвы с ненарушенной дерниной, мигрируют вглубь почвы очень медленно и, как правило, даже через несколько лет после попадания на поверхность более 90 % запаса радионуклидов сосредоточено в верхнем слое почвы толщиной 2 – 3 см [1]. Учитывая это, для оценки загрязнения радионуклидами населённых пунктов на территории 30 км зоны и других районов, пострадавших во время аварии на Чернобыльской АЭС, пробы почвы отбирались с помощью стальных колец (диаметр кольца 140 мм и толщина 50 мм), которые фиксировали площадь поверхности и толщину взятого слоя почвы. Каждая отобранная проба почвы вместе с кольцом упаковывалась в полиэтиленовый пакет и в таком виде доставлялась в лабораторию.

Традиционно перед проведением гамма-анализа пробы гомогенизируются, чтобы соблюдались идентичные условия при измерении эталона и измеряемой пробы, но такая пробоподготовка требует значительных затрат времени, специально оборудованных помещений и обученного персонала. Кроме того, при наличии «горячих» частиц механическое перемешивание не гарантирует получение гомогенной пробы. Поэтому в условиях дефицита времени в июне 1986 г. нами был предложен и реализован метод гамма-анализа проб, отобранных с помощью кольца, который исключает этап пробоподготовки.

Заранее (априори) известно, что в пробах почвы, отобранных без нарушения дернины, радионуклиды распределены неравномерно по глубине: фиксированы в виде тонкой плёнки на поверхности почвы или их содержание падает с глубиной, например по экспоненциальному закону.

Вопросы калибровки гамма-спектрометра по эффективности регистрации в случае измерения не гомогенизированных проб решаются, если реальная проба с неравномерным распределением активности по глубине заменяется идеализированной пробой, в которой активность сосредоточена в верхнем слое почвы толщиной $\delta$ и распределена в этом слое равномерно. Эквивалентная толщина $\delta$ загрязнённого слоя почвы – это такая толщина слоя почвы с равномерно распределённой активностью, при которой интенсивность излучаемого с поверхности почвы потока не рассеянных гамма-квантов (например, гамма-квантов с энергией 661,7 кэВ, обусловленных $^{137}Cs$) равна интенсивности потока квантов, испускаемого реальной пробой. Другими словами, реальная проба с не равномерным распределением активности по глубине заменяется идеализированной пробой, в которой активность сосредоточена в верхнем слое толщиной $\delta$ ($\delta \leq 50$ мм) и распределена в этом слое равномерно, причём толщина слоя такова, что гамма-излучение на поверхности реальной и идеализированной пробы имеет одинаковую интенсивность.

Чтобы определить эквивалентную толщину $\delta$, каждая проба, отобранная кольцом, измерялась дважды: первый раз верхний слой почвы обращён к детектору («травой вниз»), а второй раз проба обращена к детектору нижней поверхностью («травой вверх»). Очевидно, отношение скоростей счёта импульсов в пике полного поглощения при измерении пробы с разных сторон однозначно определяет величину $\delta$. К примеру, если активность в пробе распределена равномерно (т.е. эквивалентная толщина равна полной высоте кольца – 50 мм), то скорость счёта в выбранном пике будет одинакова при обоих измерениях. Если же активность фиксирована в тонком поверхностном слое почвы ($\delta \approx 0$), то отношение скоростей счёта при первом и втором измерении будет иметь максимальное значение. Величину отношения скорости счёта в пике полного поглощения при измерении пробы в положении «травой вниз» к скорости счёта в этом же пике при положении пробы «травой вверх» обозначим коэффициентом R.

Для определения эффективности регистрации гамма-квантов от пробы, отобранной кольцом и эквивалентной толщиной $\delta = 10$ мм, раствором $^{137}Cs$ известной удельной активности заполняли плоскую пластмассовую кювету диаметром 140 мм до высоты 10 мм. Далее эту кювету с помощью центрирующей насадки размещали на поверхности детектора, определяли скорость счёта импульсов в аппаратурном пике, соответствующем энергии 661,7 кэВ, и рассчитывали эффективность ($\varepsilon$) регистрации. Моделируя перевёрнутую пробу («травой вверх»), создавали между детектором и кюветой с раствором $^{137}Cs$ зазор толщиной 40 мм, заполненный неактивным поглотителем, в качестве которого использовали хлористый натрий (насыпная плотность поглотителя 1,3 г/см$^3$). Определив скорость счёта в пике 661,7 кэВ, рассчитывали отношение R для пробы эквивалентной толщиной 10 мм. Чтобы определить R и эффективность регистрации для пробы с эквивалентной толщиной загрязнённого слоя $\delta$ = 15 мм, кювета заполнялась радиоактивным раствором до высоты 15 мм, а при моделировании перевёрнутой пробы толщина слоя поглотителя составляла 35 мм и т.д. Посредством повторения подобных процедур для разных значений эквивалентной толщины получили зависимости $\delta$ = f(R) и $\varepsilon$ = $\theta(\delta)$.

На практике при гамма-анализе проб, отобранных кольцом, целесообразно ввести в программу расчёта активности постоянное значение эффективности $\varepsilon_0$, характеризующее эквивалентную толщину $\delta_0$, а при расчёте активности пробы с эквивалентной толщиной $\delta_i$, которой соответствует эффективность $\varepsilon_i$, вводить поправочный множитель F = $\varepsilon_0 / \varepsilon_i$. В качестве примера в табл.1 представлена полученная нами зависимость $\delta$ и F от R для детектора ДГДК–100Б, рассчитанная для $\delta_0$ = 30 мм ($R_0$ = 1,58). С помощью табл.1 находится поправочный множитель F, на который следует умножить рассчитанные на персональном компьютере значения активности (запаса) радионуклидов, полученные при измерении пробы в положении «травой к детектору». Экспериментально установлено, что зависимость поправки F от отношения R, полученная для гамма-квантов $^{137}Cs$, справедлива и для других радионуклидов ($^{134}Cs$, $^{106}Ru$, $^{144}Ce$ и др.).

При высоких уровнях радиоактивного загрязнения почвы размещать пробы почвы в геометрии «кольцо» вплотную к детектору допустимо только, если используются гамма-спектрометры способные работать при высоких загрузках спектрометрического тракта (например, спектрометры фирм Ortec, Canberra), при использовании спектрометров отечественного производства, чтобы уменьшить загрузку спектрометрического тракта, пробу удаляли от детектора на фиксированное расстояние (100 мм) с помощью специальной втулки.

Очевидно, что при значениях R близких к единице, применяемый метод отбора и анализа проб приводит к занижению запаса радионуклидов (Бк/м $^2$), поскольку в этом случае заметная доля радионуклидов находится на глубине более 5 см. Если объёмная плотность измеряемых образцов почвы сильно отличается от единицы, целесообразно ввести поправку, учитывающую разное поглощение гамма-квантов в материале пробы и в материале (матрице) калибровочного источника (в воде, если калибровка проводится с помощью образцовых радиоактивных растворов). В работе [2] показано, что эта поправка (множитель) может быть рассчитана по формуле:

$$m = \rho\mu[1 - \exp(-\rho_0\mu_0\delta)] / \rho_0\mu_0[1 - \exp(-\rho\mu\delta)] , \qquad (1)$$

Т а б л и ц а 1. Зависимость эквивалентной толщины δ и поправочного множителя F от отношения (R) скоростей счёта импульсов при положении измеряемой пробы «травой вниз» и «травой вверх».

| Отношение R | Эквивалентная толщина δ, мм | Поправка F* |
|---|---|---|
| 1,0 | 50 | 1,16 |
| 1,2 | 42 | 1,12 |
| 1,4 | 36 | 1,05 |
| 1,6 | 31 | 0,98 |
| 1,8 | 27 | 0,92 |
| 2,0 | 24 | 0,85 |
| 2,2 | 21 | 0,79 |
| 2,4 | 19 | 0,73 |
| 2,6 | 16 | 0,69 |
| 2,8 | 14 | 0,66 |
| 3,0 | 12 | 0,63 |
| 3,2 | 9 | 0,61 |
| 3,4 | 7 | 0,59 |
| 3,4 | 4 | 0,57 |

*– эффективность регистрации для δ = 30 мм принята за единицу

где  δ – эквивалентная толщина загрязнённого слоя почвы, см

μ(E) – коэффициент массового ослабления гамма-излучения для почвы, см$^2$/г;

μ$_0$(E) – то же для воды, см$^2$/г;

ρ – плотность пробы, г/см$^3$;

ρ$_0$ – плотность воды, г/см$^3$;

E – энергия гамма-квантов, кэВ.

С целью практического использования по формуле (1) была рассчитана величина поправки m для проб различной плотности и энергии гамма-квантов 661,7 кэВ. Расчёты проведены для μ$_0$(661,7) = 0,0857 см$^2$/г и μ(661,7) = 0,077 см$^2$/г. Результаты этих расчётов и данные табл.1 сведены в табл. 2, которая может быть использована не только при определении запаса в почве $^{137}$Cs, но также при расчётах запаса других радионуклидов с близкими значениями энергий гамма-квантов ($^{134}$Cs, $^{106}$Ru). Для $^{144}$Ce, испускающего гамма-кванты с энергией 133,5 кэВ, значения m следует рассчитывать по формуле (1) при $\mu_0(134) = 0,148$ см$^2$/г (в случае калибровки по эффективности с применением образцовых радиоактивных растворов) и $\mu(134) = 0,138$ см$^2$/г.

Применение описанного метода позволило при круглосуточной работе двух гамма-спектрометров в июне 1986 г. проанализировать в течение 20 дней свыше 2000 проб почвы, отобранных кольцами в населённых пунктах, расположенных в радиусе 100 км от Чернобыльской АЭС.

## СПИСОК ЛИТЕРАТУРЫ

Т а б л и ц а 2. Зависимость поправки mF от отношения R и массы пробы.

| Масса пробы (без кольца), г | Отношение скоростей счёта импульсов при измерении пробы в положении «трава вниз» и «трава вверх» (R). | | | | | | | | | | | |
|---|---|---|---|---|---|---|---|---|---|---|---|---|
| | 1,0 | 1,2 | 1,4 | 1,6 | 1,8 | 2,0 | 2,2 | 2,4 | 2,6 | 2,8 | 3,0 | 3,2 |
| 300 | 1,02 | 1,00 | 0,95 | 0,90 | 0,85 | 0,80 | 0,75 | 0,69 | 0,66 | 0,63 | 0,61 | 0,59 |
| 400 | 1,04 | 1,02 | 0,97 | 0,92 | 0,87 | 0,81 | 0,75 | 0,70 | 0,67 | 0,64 | 0,61 | 0,60 |
| 500 | 1,07 | 1,04 | 0,99 | 0,93 | 0,88 | 0,82 | 0,76 | 0,71 | 0,67 | 0,64 | 0,62 | 0,60 |
| 600 | 1,09 | 1,06 | 1,00 | 0,94 | 0,89 | 0,83 | 0,77 | 0,71 | 0,68 | 0,65 | 0,62 | 0,60 |
| 700 | 1,12 | 1,09 | 1,02 | 0,96 | 0,90 | 0,83 | 0,78 | 0,72 | 0,68 | 0,65 | 0,62 | 0,61 |
| 800 | 1,14 | 1,11 | 1,04 | 0,97 | 0,91 | 0,84 | 0,79 | 0,73 | 0,68 | 0,66 | 0,63 | 0,61 |
| 900 | 1,17 | 1,13 | 1,06 | 0,99 | 0,93 | 0,85 | 0,79 | 0,73 | 0,69 | 0,66 | 0,63 | 0,61 |
| 1000 | 1,20 | 1,15 | 1,08 | 1,00 | 0,94 | 0,86 | 0,80 | 0,74 | 0,70 | 0,67 | 0,64 | 0,61 |
| 1100 | 1,23 | 1,18 | 1,09 | 1,02 | 0,95 | 0,87 | 0,81 | 0,75 | 0,70 | 0,67 | 0,64 | 0,62 |
| 1200 | 1,25 | 1,20 | 1,11 | 1,03 | 0,96 | 0,88 | 0,82 | 0,75 | 0,71 | 0,68 | 0,64 | 0,62 |
| 1300 | 1,28 | 1,22 | 1,13 | 1,05 | 0,97 | 0,89 | 0,83 | 0,76 | 0,71 | 0,68 | 0,65 | 0,62 |

\\